# *New nodes attack strategies for real complex weighted networks*

Q. Nguyen[a,b], D. Cassi[c], M. Bellingeri [c,d]

[a]Division of Computational Mathematics and Engineering, Institute for Computational Science, Ton Duc Thang University, Ho Chi Minh City, Vietnam

[b]Faculty of Finance and Banking, Ton Duc Thang University, Ho Chi Minh City, Vietnam

[c]Dip. Scienze Matematiche, Fisiche e Informatiche, Università di Parma, Parco Area delle Scienze, 7/A, 43124 Parma

[d]Dipartimento di Fisica, Politecnico di Milano, Piazza Leonardo da Vinci 32, 20133, Milano, Italy

Corresponding author email address: nguyenquang@tdtu.edu.vn

**Abstract**

In this work we introduce a new nodes attack strategy removing nodes with highest conditional weighted betweenness centrality (CondWBet). We compare its efficacy with well-known attack strategies from literature over five real-world complex weighted networks. We use the network weighted efficiency (WEFF) like a measure encompassing the weighted structure of the network in addition to the commonly used binary-topological measure, the largest connected cluster (LCC). We find that the recently proposed conditional betweenness strategy (CondBet) (Nguyen et al. 2019) is the best to fragment the LCC in all cases. Further, we find that the introduced CondWBet strategy is the best to decrease the network efficiency (WEFF) in 3 out of 5 cases. Last, CondWBet is be the most effective strategy to reduce WEFF at the beginning of the removal process whereas the Strength that removes nodes with highest link weights first, shows the highest efficacy in the final phase of the removal process when the network is broken in many small clusters. These last outcomes would suggest that a better attacking strategy could be a combination of the CondWBet and Strength strategies.

1. Introduction

The study of real-world complex networks has attracted much attention in recent decades because a large number of real complex systems can be abstracted as networks (Costa et al. 2011). One of the fundamental research topics is their robustness (resilience), i.e. capacity of the network to hold its functioning when a proportion of nodes is removed/blocked (Albert and Barabási 2002; Cohen et al. 2000; Callaway et al. 2000; Iyer et al. 2013; Bellingeri, Agliari, and Cassi 2015; Bellingeri et al. 2014; Dall'Asta et al. 2006; Nguyen and Nguyen 2018; Wandelt et al. 2018). Previous studies showed that most real-world networks are resilient to random failure (Cohen et al. 2000) but can disintegrate quickly when a small proportion of most connected nodes (hubs) are attacked (Albert, Jeong, and Barabási 2000). Furthermore, one of the remarkable observations is when the proportion of nodes removed is high enough, a phase transition occurs and the probability for the existence of a largest connected cluster (LCC or giant cluster) in the network abruptly collapses (i.e. the network loses its global nodes connectivity). Monte-Carlo simulation is usually used to run the attack removing nodes according to a structural criterion and tracing the network damage using measures/indicators of its functioning. Overall findings showed that nodes attack strategies based on betweenness centrality are highly efficient to dismantle the LCC for most model and real-world networks (Bellingeri et al. 2014; Iyer et al, 2013; Nguyen and Nguyen 2018; Wandelt et al. 2018). The attack based on the betweenness centrality (Bet) ranks and removes nodes according to their role in routing the shortest paths in the network, i.e. nodes with higher betweenness centrality are the ones where pass the major amount of shortest paths among other nodes (Bellingeri et al. 2014; Wandelt et al. 2018). This finding indicated that the most important nodes are not necessarily the most connected ones (i.e. hubs), but they can be nodes with a medium of low connectivity level but playing pivotal role connecting different parts of the network. However, recent work showed that the Bet strategy becomes ineffective at the end of the removal process due to the inherent nature of betweenness's definition - for a fully connected network (e.g. a complete graph), all node's betweenness are zero (Nguyen et al. 2019). At this stage, the Bet strategy was unable to break the fully connected LCC for a very long time and finally, it fails to completely break the network earlier than other strategies, such as the degree based strategy. To overcome this limitation of the betweenness nodes attack strategies, (Nguyen et al. 2019) introduced the 'conditional attack strategy' (CondBet) that remove nodes according to their betweenness only if they belong to the LCC, demonstrating how the new conditional strategy outperforms the classic nodes attack strategies.

However, the aforementioned researches investigate binary complex networks where the interaction between two nodes is either 1 (have an edge) or 0 (no edge). In reality, many real-world complex networks are naturally weighted with some interactions value associated to the links. For example, in a communication network the link weights may represent the frequency of email exchanges (Leskovec, Kleinberg, and Faloutsos 2007; Yin et al. 2017); in the Internet network link weights could be the connection bandwidth magnitude (Leskovec et al. 2007; Ripeanu, Foster, and Iamnitchi 2002)Leskovec et al. 2007, Ripeanu et al. 2002); in a flights transportation network the link weights could be the total passengers flowing between airports (M. Barthélemy et al. 2005). In some case, real networks own very large link weight heterogeneity, with link weights spanning over five order of magnitude, such as the passengers flow between airports (Dall'Asta et al. 2006). This implies that to perform more real and precise network description it is necessary to account the link weights heterogeneity (Bellingeri et al. 2019; Bellingeri and Cassi 2018).

Bellingeri and Cassi (2018) analyzed the robustness of real networks to different nodes attack strategies using binary and weighted indicators of network functioning. The authors found that the inclusion of link weights in the analyses changes the network response to nodes attack, outlining the importance to investigate the performance of nodes attack strategies in weighted networks (Bellingeri and Cassi 2018). This leads us to the questions: 1) Which is the best nodes attack strategy to harm real weighted networks? and 2) How does the inclusion of link weight changes the global efficacy of the nodes attack betweenness-based strategies?

In this paper, we analyzed the robustness of a high quality dataset of real-world weighted networks from different fields of science. We implemented six attack strategies based on binary and weighted properties of the nodes. We adopted the recalculated (adaptive) version of the nodes attack strategies in which the nodes rank is updated after each nodes removal (Wandelt et al. 2018). Here we introduce for the first time the conditional nodes weighted betweenness centrality attack strategies (CondWBet) (the weighted version of the CondBet strategy) that remove nodes of highest weighted betweenness centrality inside the LCC. We test the efficacy of the different nodes attack strategies by computing their impact on network robustness using both binary and weighted measures of network functioning, i.e. the largest connected cluster (LCC), the weighted efficiency (WEFF).

We found that the conditional betweenness attack strategy (CondBet) is the best to fragment the LCC in all the networks confirming previous evidences (Nguyen et al. 2019).

Then, when the goal is to decrease the network efficiency measure (WEFF), in 3 out of 5 networks the weighted conditional betweenness strategy (CondWBet) is the best strategy. On the other hand, in other two networks the nodes strength attack strategies (Strength, that removes nodes with highest sum of link weights) and the binary conditional betweenness strategy (CondBet) are the best strategy.

- **Strength**: Strength attack strategy removes nodes in decreasing rank of strength. The strength of the node is the sum of the link weights pointing to it (Bellingeri and Cassi 2018). The strength of the node it is defined as the 'weighted degree' and for this reason the Strength can be viewed as the weighted counterpart of the Deg strategy.
- **WBet**: Weighted Betweenness attack strategy removes nodes according to their weighted betweenness centrality. Like the binary version of the nodes betweenness centrality, the weighted betweenness centrality for each node is the number of the shortest paths that pass through the nodes. The difference in this case is that the shortest paths among nodes are weighted, i.e. each link contributes to the shortest path according to their inverse weight. This is a standard procedure making links of higher weight equivalent to 'larger, faster and shorter' route' between nodes. The weighted shortest path between two nodes is the path minimizing the sum of the inverse link weights between that pair of nodes (Brandes 2001).
- **CondWBet**: Conditional weighted betweenness attack strategy is introduced for the first time in this work. It is the weighted version of the CondBet. The CondWBet strategy removes nodes inside the LCC according to their recalculated weighted betweenness centrality.

**The real-world networks**

We analyzed the efficacy of the nodes attack strategies on five real-world weighted networks (Table 1). The first two are financial networks constructed from stocks in the SP500 index in the US market using the threshold method (Boginski, Butenko, and Pardalos 2005) at different time periods. We keep the value of the correlation coefficients that are higher than the threshold as link weights. By adjusting the threshold, we can obtain two networks with a similar average degree and number of nodes but with different topological structures. The first network, (i) the SP500_1 is built from Feb. 1993 to Feb. 1996 is relatively uniformly connected and contains 315 nodes and 8706 links (see Figure 6 a) while the second (ii) SP500_2 is built from May 1999 to May 2002 contains several well-connected clusters connecting to the central bulk through intermediated stocks and contains 371

The paper is organized as follows. In Section 2, we describe the data and methods used in this work. Section 3 presents the empirical findings: the efficacy of six intentional attacking strategies over five real-world weighted networks. Finally, Section 4 summarizes the results and concludes.

**2. Method and data**

**The nodes attack strategies**

We simulate nodes attacks on the networks using strategies belonging to two main groups, the binary and the weighted-based strategies. In each group, we use node degree, node betweenness centrality and the conditional betweenness centrality for nodes ranking. In case of nodes ranking ties (e.g. nodes with equals rank, we randomly sort one of them).

In total, we adopt six nodes attack strategies:

- **Deg**: Degree based attack strategy removes nodes according to their degree, i.e. the degree of the node is the number of links to it (Albert and Barabasi 2002, Bellingeri et al. 2014, Wandelt et al. 2018). This is the simple and the oldest type of node attack widely used to test the networks robustness.

- **Bet**: Betweenness attack strategy removes the nodes with the highest betweenness centrality first. The betweenness centrality of the node is a macro-scale network metric measuring the number of times a node appears in the shortest path between all pairs of nodes in the network (Bellingeri et al. 2014; Brandes 2001; Wandelt et al. 2018).

- **CondBet**: Conditional betweenness attack strategy is the improved version of the Bet (Nguyen et al. 2019). The CondBet removes the node with highest betweenness if it is in the LCC, otherwise it removes the node with highest betweenness inside the LCC. In other word, the CondBet removes nodes inside the LCC only, maximizing the efficacy of the Bet strategy into disrupt the LCC. It has been shown that the CondBet performs better than Bet and it owns higher efficacy to fragment the LCC on a variety of real-world networks (Nguyen et al. 2019).

nodes and 10636 links (see Figure 6 c) (Nguyen et al. 2019). Other three networks are: (iii) the co-authorship undirected network of scientists working on network theory and experiments (NetScience) compiled by Newman (2003). Nodes represent authors and link weights represents the number of common papers. The network includes 1589 nodes and 2742 links. (see Figure 6 e) (iv) the network of 6005 nodes-peoples who trade using Bitcoin on a platform called Bitcoin OTC (Bitcoin) (Kumar et al. 2017, 2018). The weight of the links represents the rate of members on other members in a scale of -10 (total distrust) to +10 (total trust) in steps of 1. We simplified the network by removing self-node links and converting to an indirect network. It results in a network of 6005 nodes and 21492 links. (see Figure 6 g) (v) the network of 500 busiest airports in U.S. (Colizza, Pastor-Satorras, and Vespignani 2007) where nodes represents airports and the weight of a link identifies the normalized passengers flowing between two airports/nodes (see Figure 6 i) (Dall'Asta et al. 2006)

Table 1:

| Name | SP500_1 | SP500_2 | NetScience | Bitcoin | USAirport |
|---|---|---|---|---|---|
| Type | Financial | Financial | Co-author | Social | Traffic |
| Nodes | Stock | Stock | Scientist | Member | Airport |
| Weight | Correlation coefficient | Correlation coefficient | Number of common papers | Trusting rate | Passengers flow |
| $N$ | 315 | 317 | 1589 | 6005 | 499 |
| $L$ | 8706 | 10636 | 2742 | 21492 | 2980 |
| $<k>$ | 27.6 | 28.6 | 1.72 | 3.57 | 5.97 |
| $<s>$ | 10.349 | 18.817 | 1.497 | 6.538 | 0.0675 |
| $<w>$ | 0.187 | 0.328 | 0.433 | 0.913 | 0.018 |
| Ref. | Nguyen et al. 2019 | Nguyen et al. 2019 | Newman 2003; Boccaletti et al. 2006 | Kumar et al. 2017, 2018 | Colizza et al. 2007 |

*Structural statistics of the real-world networks: network type, nodes type, weight definition, N nodes' number, L links' number, <k> average node degree, <s> average node strength, <w> average link weight and Reference's source*

**The network robustness measures**

We used two measures of network robustness along the nodes attack process, e.g. the size of the largest connected cluster (LCC) and the weighted efficiency (WEFF). The LCC is a widely used measure of the network robustness (also known as largest connected component or giant component) and it accounts the highest number of connected nodes in the network (Albert and Barabási 2002; Iyer et al. 2013; Bellingeri et al. 2014; Bellingeri et al. 2015; Morone and Makse 2015; Wandelt et al. 2018). The LCC is a simple indicator evaluating the binary-topological connectivity of the network nodes and we use it like a binary measure of the network functioning not considering the underlying link weights structure.

The network efficiency (WEFF) is introduced by (Latora and Marchiori 2001) with the goal to account the network information delivery rate in the network. WEFF is the sum of the inverse of the weighted shortest paths among nodes (Latora and Marchiori 2001):

$$Eff = \frac{1}{N \cdot (N-1)} \sum_{i \neq j \in G} \frac{1}{d(i,j)} \quad (1)$$

where *N* is the total number of nodes of network G and *d(i,j)* is the weighted shortest path between node *i* and node *j*.

The WEFF is a measure that considers the difference in link weights in the evaluation of the weighted network functioning (integrity) and can be viewed like an indicator of how efficiently the network nodes exchange information (Crucitti et al. 2003; V.Latora and M.Marchiori 2001). Recently, the network efficiency has been used to evaluate and compare the efficacy of different nodes weighted attack strategy into decrease the robustness of complex weighted networks (Bellingeri and Cassi 2018; Bellingeri et al. 2019; M. Bellingeri et al. 2020)

In addition, for each attack strategy, we compute a single value defined as the network robustness (*R*) as done in (Bellingeri et al. 2019). The value of *R* corresponds to the area below the curve of the system functioning indicators (LCC and WEFF) against the fraction of nodes-links removed.

## 3. Results

**Efficacy of the attack strategies with LCC**

We found that the CondBet is the best strategy to minimize $R_{LCC}$ for all the five networks (Figure 1 and Table 2). This outcome confirms the highest efficacy of this strategy into fragment the LCC (Nguyen et al. 2019). CondWBet is the second best strategies to fragment the LCC outperforming the classic WBet counterpart in all the network (Figure 1). We can see that at the end of the removal process, the nodes betweenness centrality such as Bet and WBet become inefficacy, especially in the SP500_2 and NetScience networks (Figure 2). The Bet and WBet nodes attack strategies get stuck in a fully connected LCC and are not able to break it for a long fraction of removals. In fact, both binary (Bet) and weighted (WBet) nodes betweenness centrality inside a fully connected LCC are zero for all the nodes (because all the paths inside this sub-network are shortest paths); thus the Bet and WBet strategies are not able to select nodes and to fragment the LCC. In Figure 2 we compare the LCC size after removing a fraction $q$ of nodes by the attack strategies outlining that the conditional strategies CondBet and CondWBet as the best to attack the LCC, this because they are able to select the most important nodes during the entire removal process.

Moreover, we find that including link weights as done in the Strength, WBet and CondWBet worsen the efficacy of the strategies with respect to their corresponding non-weighted strategy counterpart (Deg, Bet and CondBet, respectively) into reducing the size of the LCC (Figs 1 and 2). For example, in the SP500_2 network the removal of $q = 4\%$ nodes by CondBet strategy triggers the faster network network fragmentation with respect the same nodes removal fraction performed by the CondWBet weighted counterpart strategy (Fig. 3). In Figure 3 we can see that he CondBet strategy is able to fragment the network isolating a large cluster, thus producing a sharper LCC decrease. This would suggest that for the networks analyzed here adding information about link weights may degenerate the efficacy of the attack strategies to select important nodes supporting the simple binary-topological connectivity (measured by the LCC). In fact, the link weights structure may induce the weighted attack strategies to remove nodes, hence important for network functioning, not playing a major role in shaping the topological network connectivity.

**Efficacy of the attack strategies with WEFF**

When the goal is to minimize to minimize $R_{WEFF}$, the CondWBet strategy is the best strategy in 3 out of 5 networks (Figure 2, Table 2). The CondWBet is highly effective to reduce the network efficiency (WEFF) because is able to remove the nodes where passing the most of the weighted shortest paths in the network playing the major role to enhance the network efficiency WEFF (Eq. 1). For this reason, when removing the nodes with highest weighted betwenness we trigger the disruption of the weighted shortest paths in the network with a quick WEFF decrease. Furthermore, CondWBet is more efficacy than the classic betweenness strategy WBet for all networks. This can be explained by the ability of CondWBet to select nodes only inside the LCC at the end of the removal process when it is fully connected, thus providing a further efficacy with respect classic WBet.

| Network | Best strategy to reduce $R_{LCC}$ | Best strategy to reduce $R_{WEFF}$ |
|---|---|---|
| SP500_1 | CondBet | CondWBet |
| SP500_2 | CondBet | CondBet |
| NetScience | CondBet | Strength |
| Bitcoin | CondBet | CondWBet |
| USAirport | CondBet | CondWBet |

*Table 2: The most efficacy nodes attack strategy for the five real weighted networks.*

For the SP500_2 the efficacy difference between CondBet and CondWBet to reduce WEFF is very narrow, with almost negligible advantage of the CondWBet strategy (Figs 1 and 2). For this reason, we can consider CondBet and CondWBet roughly equally performing, with negligible difference due to some less important effect.

Differently, in the Netscience network the Strength is clearly the best strategy to decrease WEFF. The highest efficacy of the Strength strategy for the Netscience network can be explained by the peculiar embedding of the highest weight links (strong links). The Netscience is the social network of co-authorship, where nodes are scholars and links weights indicate the number of common papers. In this network, the strong links occur between senior and most prolific scholars leading different research groups, i.e. the strong links act as bridges between different research communities (Pan and Saramaki 2012, Bellingeri et al. 2020). The nodes of higher strength are these senior scholars publishing many papers and holding the collaborations with different research groups. For this reason, when removing nodes of higher strength in the Netscience network, we remove strong links (playing a major role into shaping WEFF) and at the same time we delete the bridge links between communities, producing an abrupt collapses of the network efficiency WEFF (Figure 4).

Last, we find an interesting 'efficacy transition' between CondBet/CondWBet and Strength strategies for all the networks, i.e. at the beginning of the removal process CondBet/CondWBet are the best strategies whereas at the end of the removal process Strength strategy shows the highest efficacy into reduce WEFF (Figure 5). This may be explained by the fact that when the network is broken apart in many small clusters, the nodes betwenness based strategies would become ineffective to produce a significant further network fragmentation. For this reason, at this stage to select highest strength nodes can intercept the remaining strong links that play the major contribution in shaping the network efficiency (WEFF). For example, the USairports network is mostly broken and contains a large number of small clusters as soon as *q = 7%* (Figure 5). At this stage the betweenness based strategies CondBet/CondWBet lose their global ability to intercept nodes bridge and the Strength strategy that remove nodes with highest link weights becomes the most efficacy strategy for reducing WEFF. Thus, before the transition the CondBet/CondWBet are the most efficacy strategy and after this transition when the network is broken into many small cluster, the Strength strategy will be the best one for the remaining removal process. Therefore, we can infer a general result for which when the network is broken enough, the Strength strategy may be one of the best strategy to decrease the network efficiency (WEFF). The transition points values are summarised in Table 3 and the topological image each networks at this points are shown in Figure 6. Further observations that are specific to each real-world network are presented in the Appendix.

|  | SP500_1 | SP500_2 | NetScience | Bitcoin | USAirport |
|---|---|---|---|---|---|
| Approximate range of the fraction of removed nodes $q$ when the transition happens | 0.67 | 0.66 | 0.04 | 0.12 | 0.07 |
| Overall highest efficacy strategy for reducing $R_{WEFF}$ | CondWBet | CondBet | Strength | CondWBet | CondWBet |

**Table 3**: Fraction of removed nodes $q$ range corresponding to the transition when the Strength strategy become more efficacy than other strategies. The overall highest efficacy strategy with minimum $R_{WEFF}$ for each real-world networks are shown in the last row. See Figures 3-4 for more details.

## 4. Conclusion

In this work, we studied the efficacy of different nodes attack strategies on real-world complex networks adopting different measures of the network functioning, both accounting the binary-topological connectedness and the link weights structure of the network. We used both classic topological-binary attacks and introducing new attacks based on weighted properties of the nodes. To develop the nodes attack strategy able to produce the fastest LCC disruption is a fundamental problem of theoretical and applied network science with huge efforts made in recent years (Albert and Barabási 2002; Albert et al. 2000; Bellingeri et al. 2014; Holme et al. 2002; Iyer et al. 2013; Morone and Makse 2015; Wandelt et al. 2018) Here, we find that the recently introduced conditional attack strategy (CondBet) outperforms the other strategy to break the LCC in all the 5 networks, confirming the highest effectiveness of this strategy (Nguyen et al. 2019). These outcomes show that conditional nodes attack strategies CondBet (Nguyen et al. 2019) may be the best methods to fragment the LCC. Furthermore, the inclusion of link weighs into nodes attack strategy results in a lower efficacy when the goal is to reduce the LCC. This would suggest that adding information about link weights may degenerate the efficacy of the attack strategies to select important nodes supporting the simple binary-topological connectivity (LCC).

Secondly, when measuring the network functioning with the network efficiency WEFF, we find that in 3 out of 5 networks, the conditional weighted betweenness strategy CondWBet outperform all the others strategy, showing that with the aim to select most important nodes in real networks it is necessary link weights. Last, we found that for all the networks the CondBet/CondWBet is the most efficient to decrease WEFF in the first phase of the removal process whereas the Strength shows the highest efficacy in the final phase. This would demonstrate that when the network is broken into many small cluster, the Strength strategy may be the best one to decrease the network efficiency (WEFF).

These last outcomes would suggest that a better attacking strategy for real-world weighted networks could be a combination of CondBet/CondWbet and Strength. Our work may help to design best the attack strategy, or inversely to design a more robust weighted network structure in practice.

## 5. Acknowledgement

This work is supported by the Vietnam National University Ho Chi Minh City (VNU-HCM), Ho Chi Minh city, Vietnam (under grant number B2017-42-01) and by the Vietnam's Ministry of Science and Technology (MOST) under the Vietnam-Italy scientific and technological cooperation program for the period of 2020-2022. Many thanks to Dr. Stefano Poletti and Prof. Francesco Scotognella for useful comments on the paper.

## 6. Author contribution statement

Please provide a statement at the end of the paper (after the Acknowledgments) QN and MB conceived the analyses, QN performed the simulation. QN, MB and DC wrote the paper.

**FIGURES**

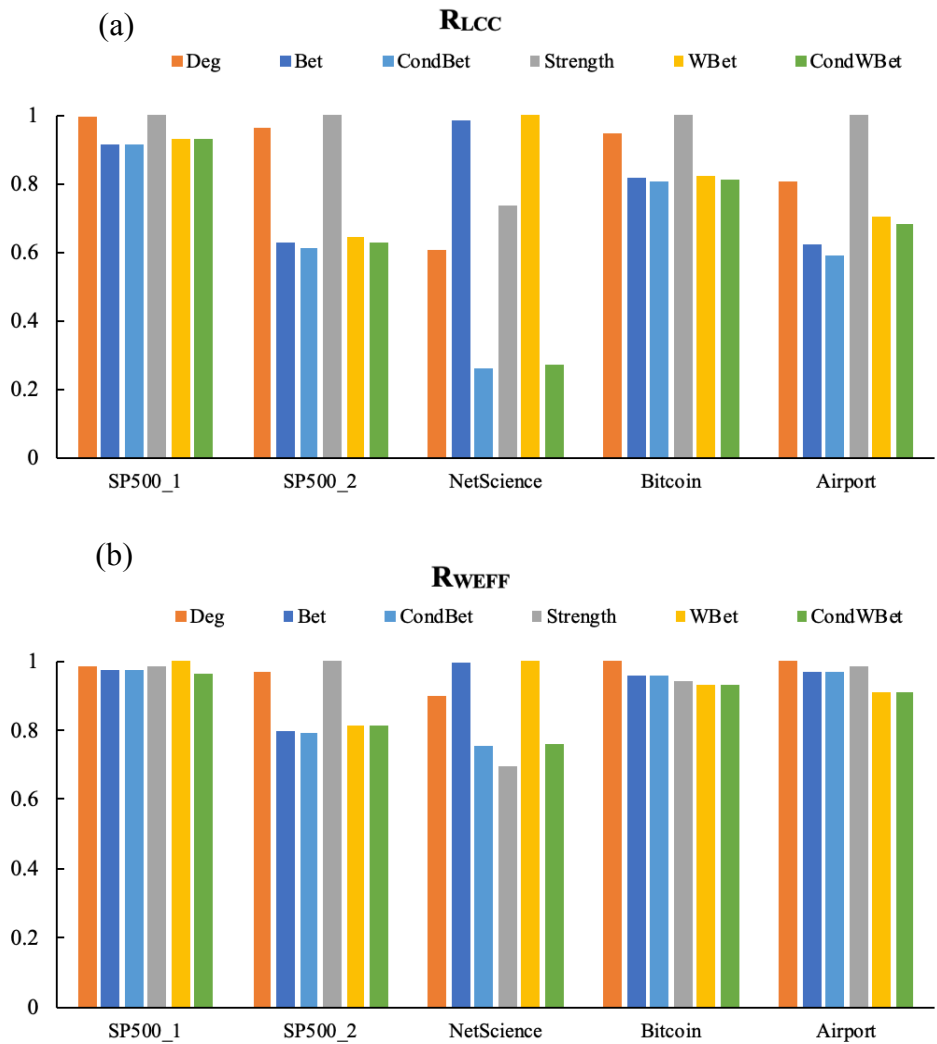

*Figure 1: The single value network robustness (R) for LCC (a) and WEFF (b) for each networks and strategies. To easily compare the efficacy of the strategies we normalized R by the max robustness value produced by a strategy for that network.*

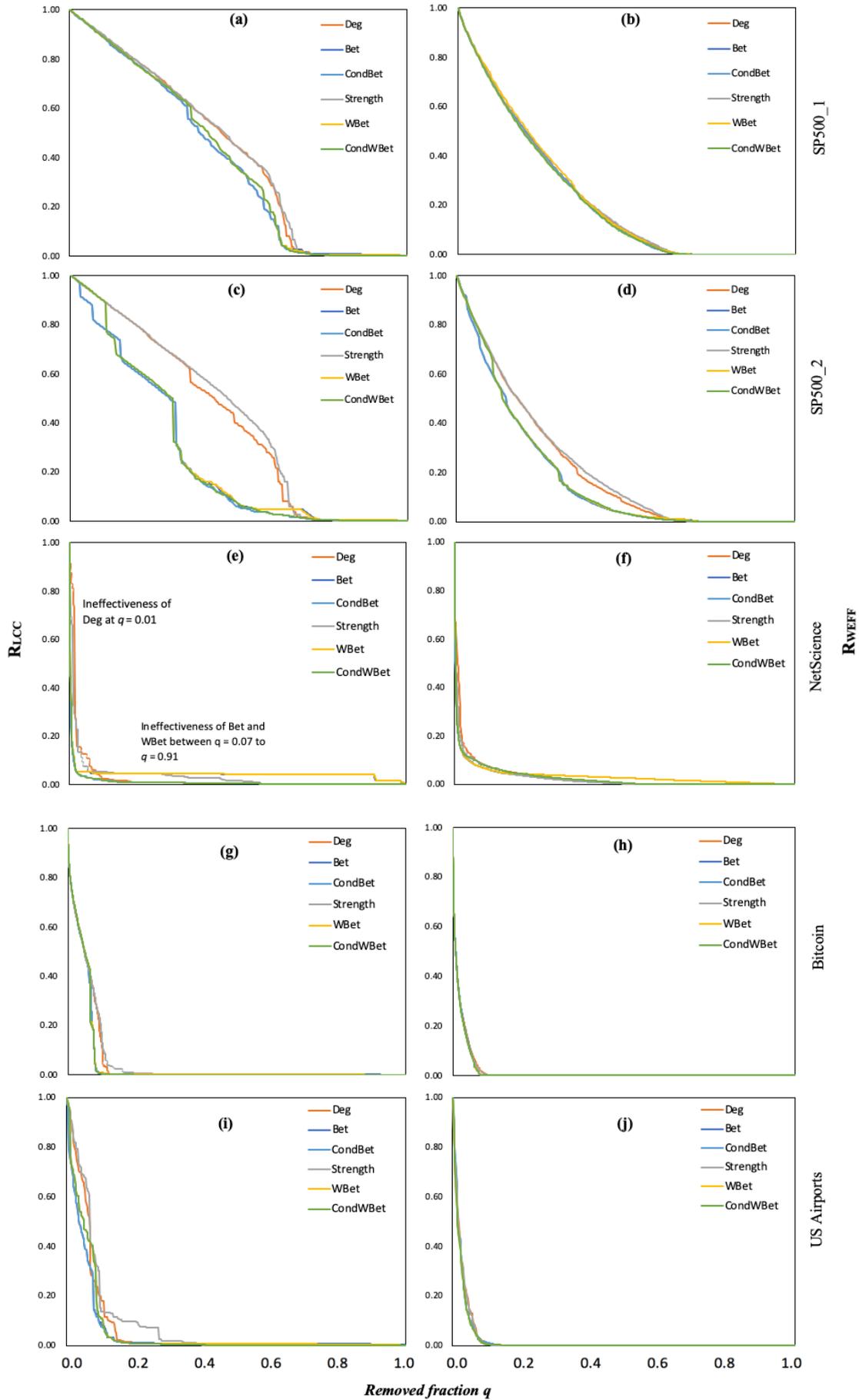

*Figure 2: Normalized largest connected cluster (LCC) and weighted efficiency (WEFF) as a function of the fraction of removed nodes (q) under different attack strategies: (a,b) the SP500 threshold network from Feb. 1993 to Feb. 1996 (SP500_1), (c,d) the SP500 threshold network from May 1999 to May 2002 (SP500_2), (e,f) the co-authors network (NetScience), (g,h) the Bitcoin trust network and (i,j) the USAirport network*

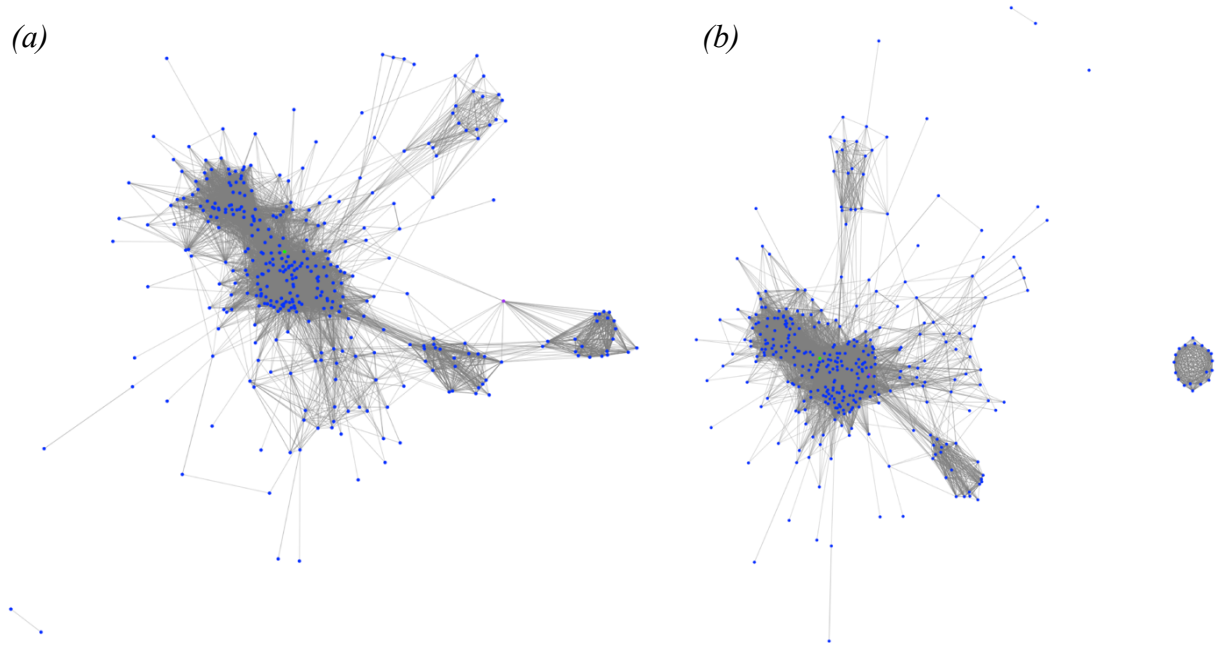

*Figure 3: Topological structure of the SP500_2 network at q = 4% attacked by a) the CondWBet strategy and b) the CondBet strategy. The CondBet strategy is more effective to fragment the network isolating a large cluster, thus producing a sharper LCC decrease than the weighted counterpart CondWBet.*

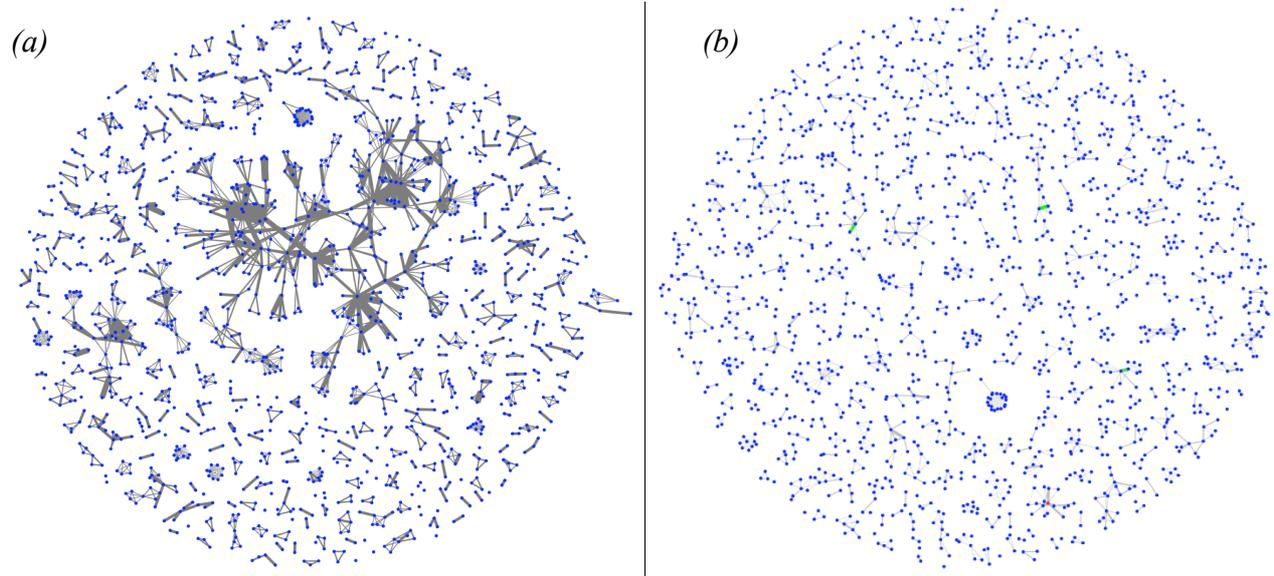

*Figure 4: Topological structure of the Netscience network at a) the beginning and b) following the q = 2% of node removal by Strength strategy. The thicker links in the left panel indicate the links of highest weight (strong links representing higher number of common papers) occurring between most prolific scholars. The removal of the 2% of the higher strength nodes produces both the network fragmentation and the deletion of the strong links thus inducing an abrupt WEFF decrease.*

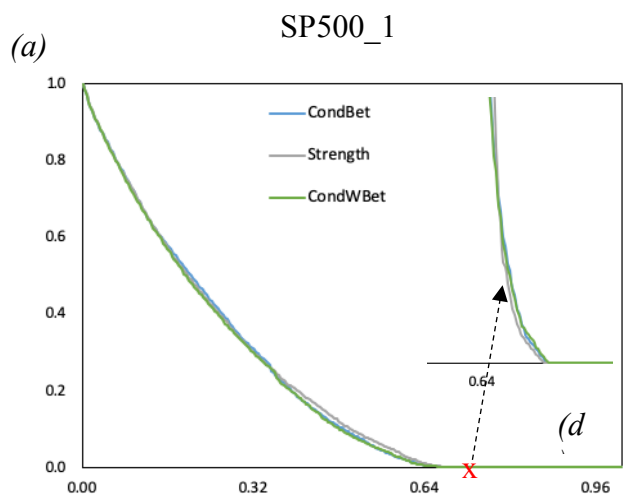
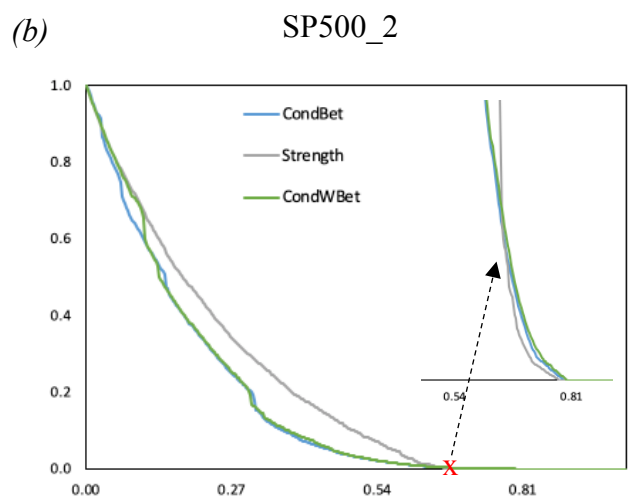
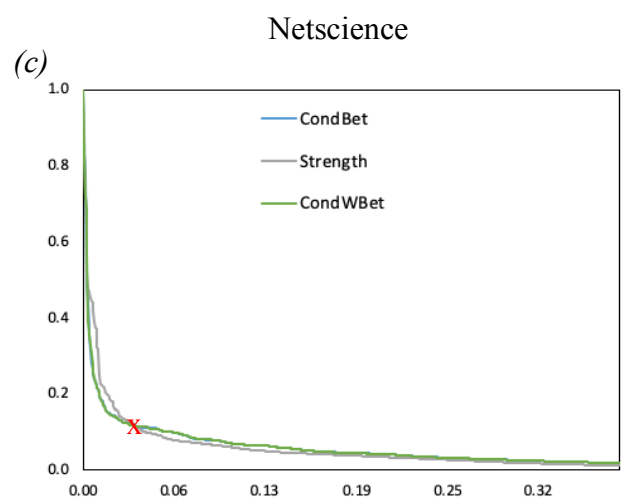
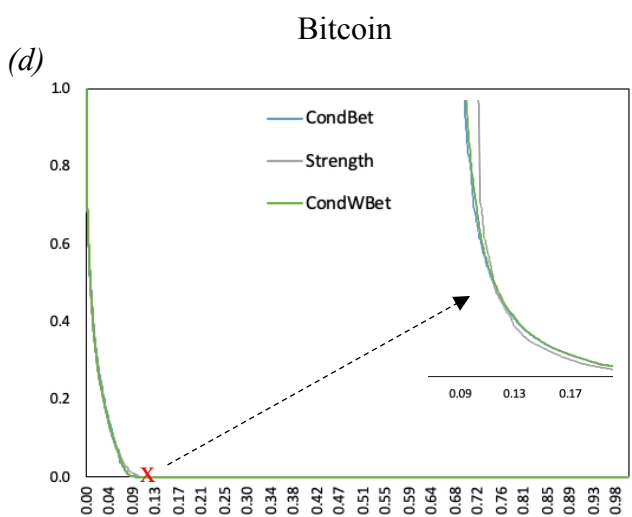
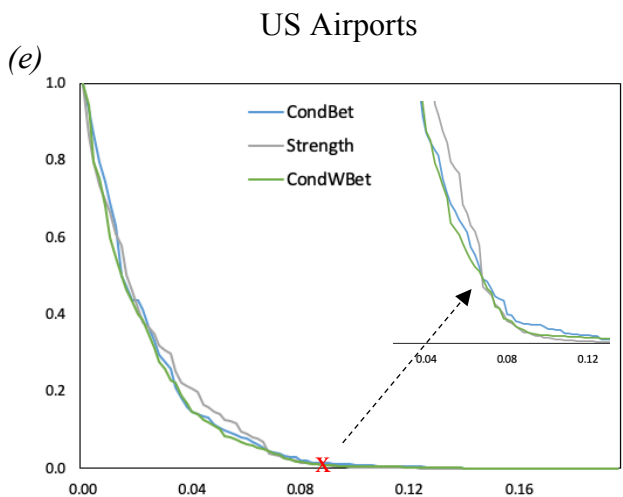

*Figure 5: The weighted efficiency (WEFF) as a function of the fraction of removed nodes (q) under three most efficacy attack strategies: Strength, CondBet and CondWBet. (a) SP500_1, (b) SP500_2, (c) NetScience, (d) Bitcoin and (e) US Airports. The transition between phases are marked by the red crosses. The inserts show a magnification of the transition (except for Netscience where this transition is clear).*

**Removed fraction q**

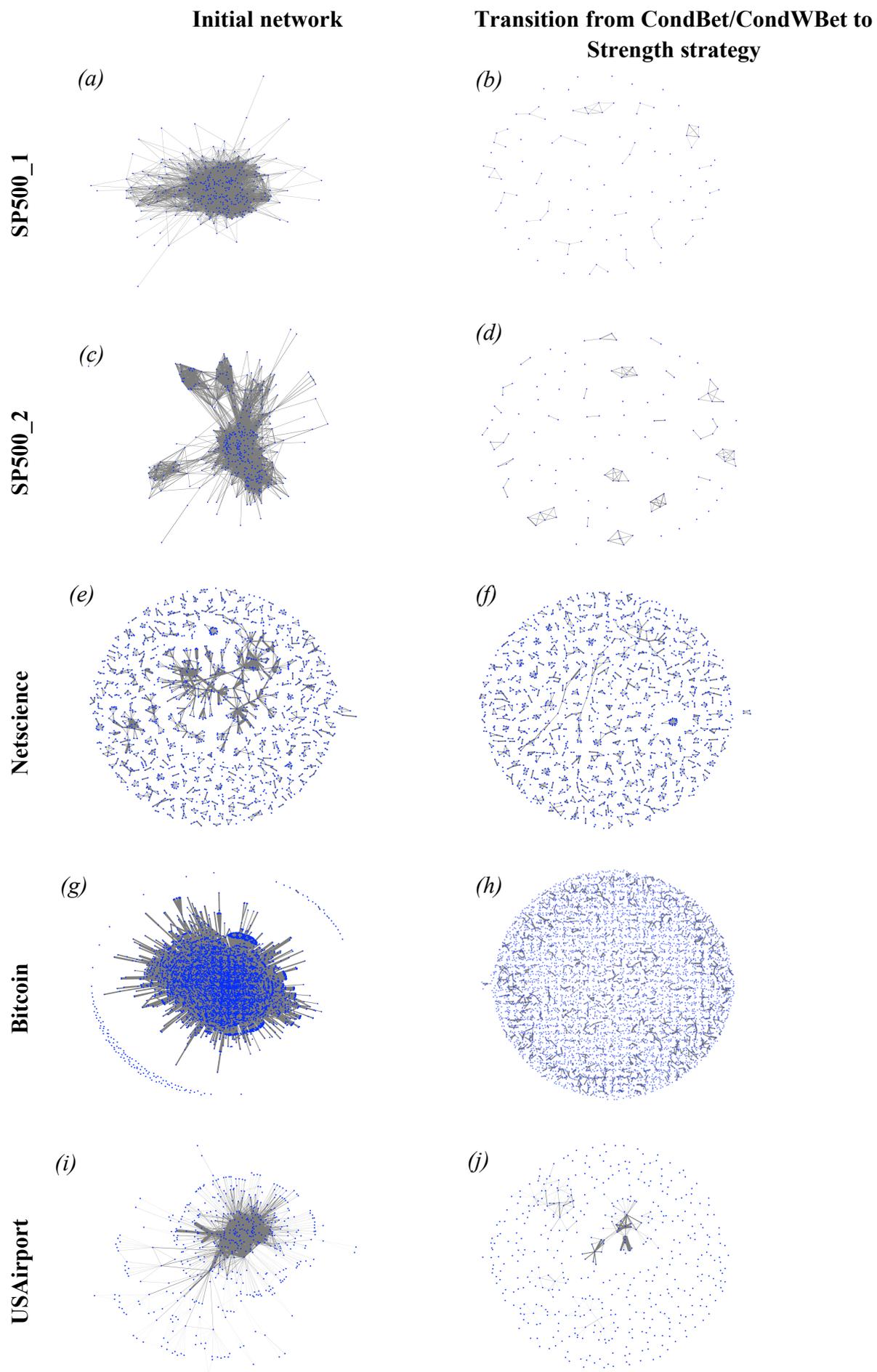

Figure 6: The topological image of the each networks in the beginning of the removal process where the conditional betweenness strategies CondBet/CondWBet are the most efficacy strategies (left column); at the

*transition point when the network are broken in many small clusters and the Strength strategy becomes the most efficacy for the remaining nodes removal process (right column).*

# APPENDIX

In this Appendix we discuss additional observations that are specific to the real-world networks, as a complement to the Result section.

**Financial networks (SP500_1 and SP500_2)**

All strategies seem relatively inefficacy to break the LCC network SP500_1 in relative to other networks. The size of the LCC only decrease linearly with the fraction of nodes removed $q$, i.e. the strategies are not able to break apart the networks and the LCCs linearly decrease as the effect of the node removal itself. Only until $q = 35\%$ then the conditional betweenness strategies (CondBet and CondWBet) can have some efficacy producing the disconnection of network nodes from the LCC (Figure 4b) and now they are able to lay below the linear decrease of the LCC (Figure 2a). Only until $q = 63\%$ then the conditional strategies can reduce the LCC to 10% of its initial size (Figure 7a). Thus in the financial network SP500_1, the CondBet and CondWBet are clearly most efficacy than all other strategies with significantly better performance to break the LCC than the related Bet and WBet strategies. The difficulty of the nodes attack strategies into break apart the LCC can be due to the fact that the network is highly and relatively uniformly connected around the main bulk (Figure 6a).

In SP500_2, all strategies perform better than in SP500_1, except for the degree-based strategies (Figure 8b). The reason can be seen from the topological image of this network as it shows higher communities effect (in Figure 6c). This structure has two-fold consequences: the betweenness based strategies are efficacy because they remove the pivotal nodes connecting different communities first, thus breaking the network apart and reducing the LCC quickly. Differently, the degree-based strategies (Deg and Strength) are inefficient because high degree nodes are usually inside a dense community and their removal does not trigger the fragmentation of the network in different clusters. This can be seen from the scattered plot of the initial betweenness *vs* degree of this network in Figure 10b. We found that in the SP500_2 there are groups of nodes with medium degree but very high betweenness. Those nodes playing the role of connecting the network are quickly removed by the betweenness-based strategy but they are not selected in principle by the degree-based strategies. Consequently, these 'higher betweenness-medium degree nodes' playing a major role in connecting the network are secondarily removed during the degree based attack process producing a slower LCC decrease. We show the higher betweenness-medium degree nodes owing a pivotal role in connecting the network in Figure 7a and b.

We found a small difference in the efficacy of the nodes attack strategies into decrease WEFF in the SP500_1 financial network (Figure 2b). Nonetheless, the CondWBet are the most efficacy to decrease WEFF until when $q = 67\%$ then the Strength is more efficace (Figure 5a). Similar observation was found with SP500_2 (at *q = 67%* as shown in table 3).

**NetScience network**

We found all the attack strategies very efficacy to reduce the LCC. This higher efficacy of the nodes attack strategies can be explained by the fact that the NetScience is the least connected network among the four studying networks. It has been shown that highest linkage density level positively affects the robustness of the networks connectivity, i.e. networks with a higher number of link per node experienced a slower decrease of the LCC when subjected to nodes removal (Albert and Barabasi 2002). The NetScience network owns average degree $<k>=1.72$ equals to one-fifteen of the average degree of the two financial networks (Table 1) and the removal of a small fraction of nodes is able to trigger the network fragmentation with faster LCC decrease.

Again, we found the conditional strategies outperform all the other strategies for reducing the LCC. Especially, when the Bet and WBet become inactive for a very long time, as soon as $q = 7\%$ until when $q = 91\%$, the Bet and WBet strategies are able to break the largest LCC. This is an interesting situation because at $q = 7\%$, a complete (fully connected) graph whose size is 16 become the LCC (Figure 8 a). As the remaining network still have 1477 nodes, the Bet and WBet will completely ignore the LCC until when all the clusters which are not fully connected are removed (Figure 8 b). The situation is different for the Deg and Strength strategies: at *q = 1%*, the same 16-node cluster already contains nodes with the highest degree and is broken by the Deg and Strength, although it is not the LCC. This is why the Deg and Strength strategies are ineffective at the early stage as shown in Figure 2 e.

For the weighted efficiency measure WEFF, we found that for this Netscience network, the Strength strategy is the most efficacy as soon as $q = 4\%$ (Figure 5c). In consequence, the Strength strategy is the most efficacy and significantly better than the second best CondBet one overall.

**Bitcoin network**

Similarly to the NetScience network, all strategies perform well into decrease the LCC (Figure 2 g). It is also because the Bitcoin network is sparse (low number of links per node) with an average degree $<k>=3.58$. We also found the inefficacy of the Bet and WBet strategies when the network is enough broken (however to a lesser extent than the NetScience). At this later stage of the removal process the Bet and WBet are not able to intercept nodes into the LCC in the network presenting a fully connected LCC and many clusters, and as a consequence, the conditional strategies CondBet, that always remove nodes inside the LCC, is the most efficient ones to vanish it.

For the weighted efficiency measure WEFF, similarly to the Netscience, the second phase starts as soon as $q = 12\%$ (Figure 5d). However, because of large gain in efficacy got in the first phase by the CondWBet strategy, the overal best strategy is still the CondWBet strategy.

**USAirport network**

All strategies perform well into decrease the LCC, in particular, the betweenness based strategy Bet and CondBet strategies outperform the others significantly (Figure 1b and 2 i).
Again, the CondBet strategy can break the complete graph at the end of the removal process, resulting in a higher efficacy than the Bet strategy. We found that for this network, the inclusion of weights into strateges worsen the efficacy in reducing the LCC for all the three strategies: Strength, WBet and CondWBet (Figure 1a and 2i). This can be seen from the scattered plot of the binary betweenness *vs* weighted betweenness of this network in Figure 9 d where one find that their correlation is the worst among networks. This may come from the fact that high binary betweenness node/airport may represent a geographically important transit point. Meanwhile, when counting the passenger flow for weighted betweenness, other airport although less well located, can still be higher ranked in terms of passenger transit.

For the weighted efficiency measure WEFF, both degree-based trategy Deg and Strength improve significantly. This suggests that high weight nodes may have important contribution in the efficiency measure of this network. However, CondWBet is still slightly better than the Strength for the whole removal process for reducing WEFF. As shown in the Figure 6 i, the USAirport network has a high degree of communities and therefore, the CondWBet works efficiently. In another word, the US 500 airports network would be most affected if the top 7% (35 airports) of the most transit point are closed. Then from this point, the whole network is almost broken (Figure 6 j) and the Strength strategy based on the total number of passenger flow will become the most efficace again.

(a)

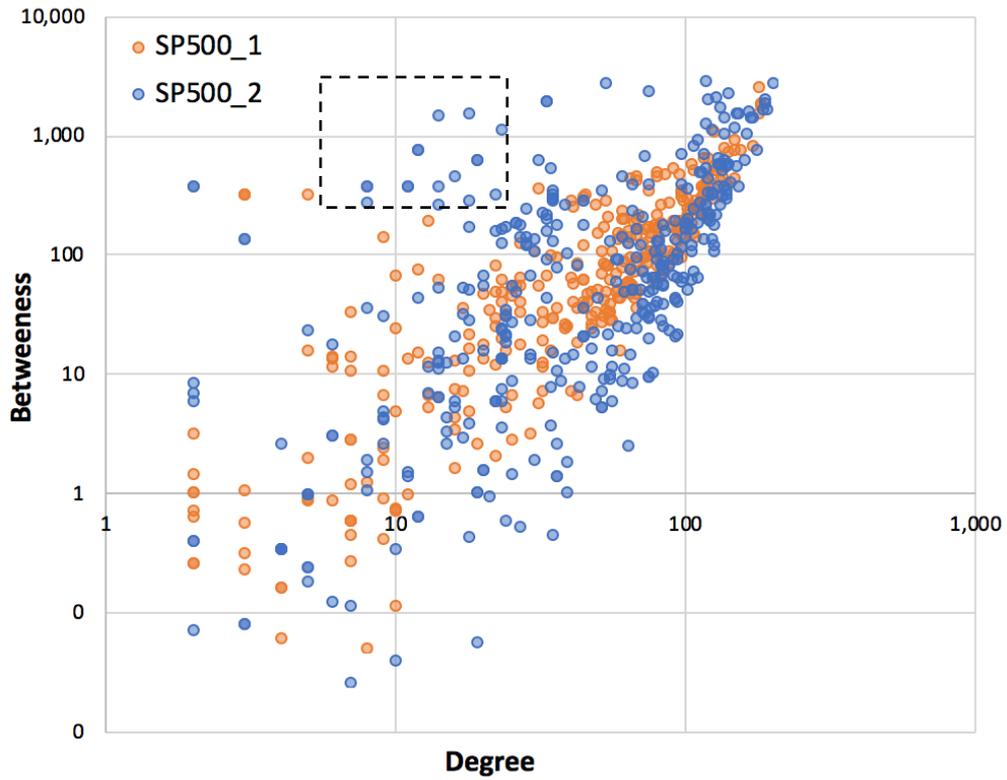

(b)

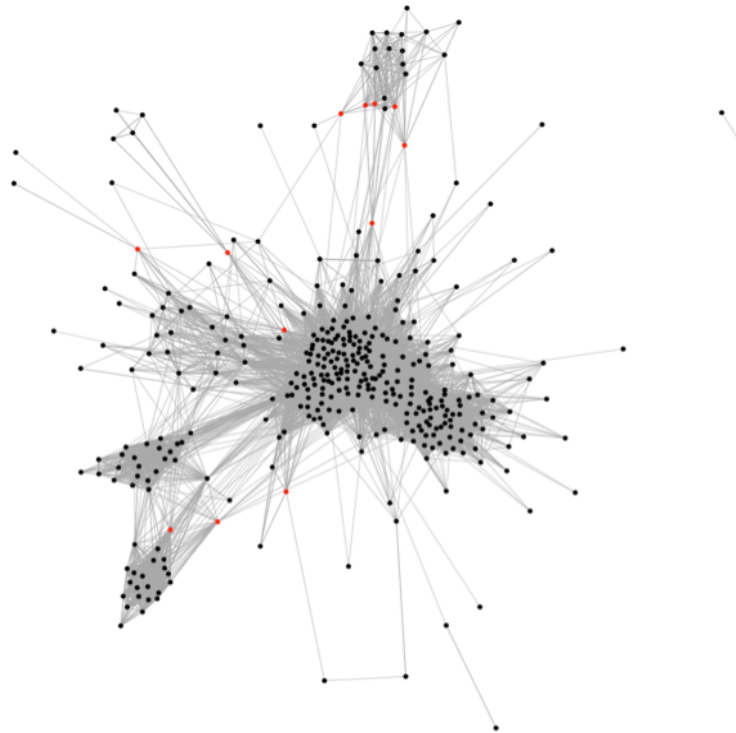

*Figure 9: (a) Binary betweenness vs. degree of the two financial networks, the SP500_1 and SP500_2 where in the SP500_2 we see a group of high betweenness-medium degree nodes of the SP500_1 (highlighted by the rectangle). (b) Those nodes are colored red in the corresponding topological graph and appear to be bridge points of the network.*

(a)

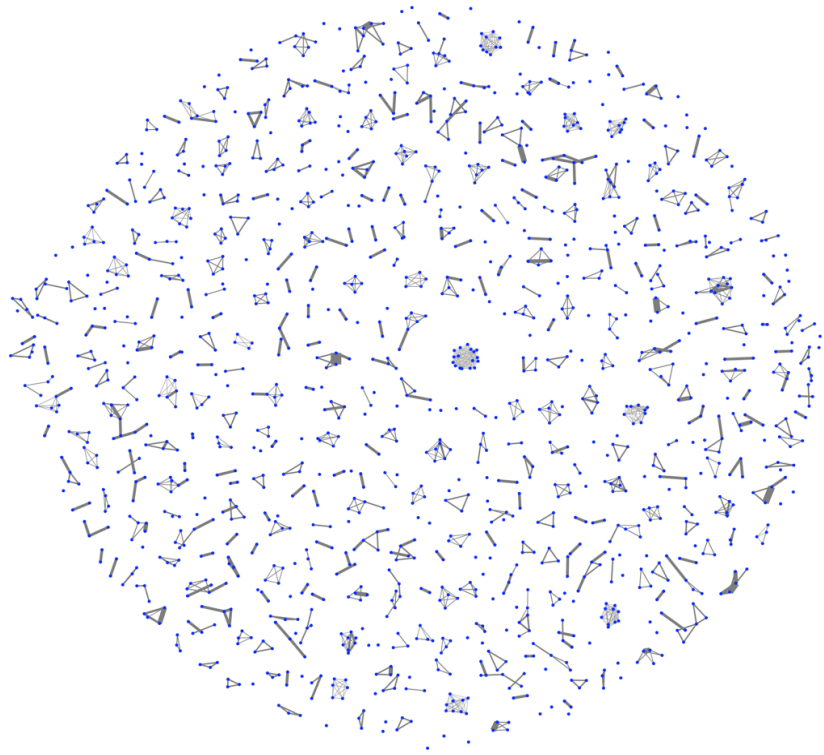

(b)

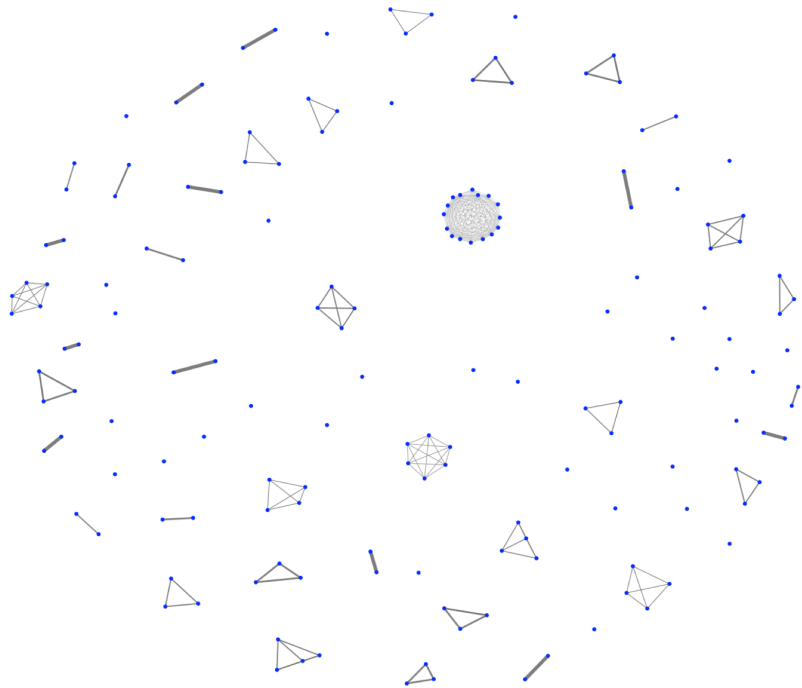

*Figure 10: The NetScience network attacked by the Bet strategy at (a) q = 7% when the LCC is a large complete graph of 16 nodes and the Bet strategy will ignore it until at (b) q = 91% when the remaining network only contain individual nodes or complete graph.*

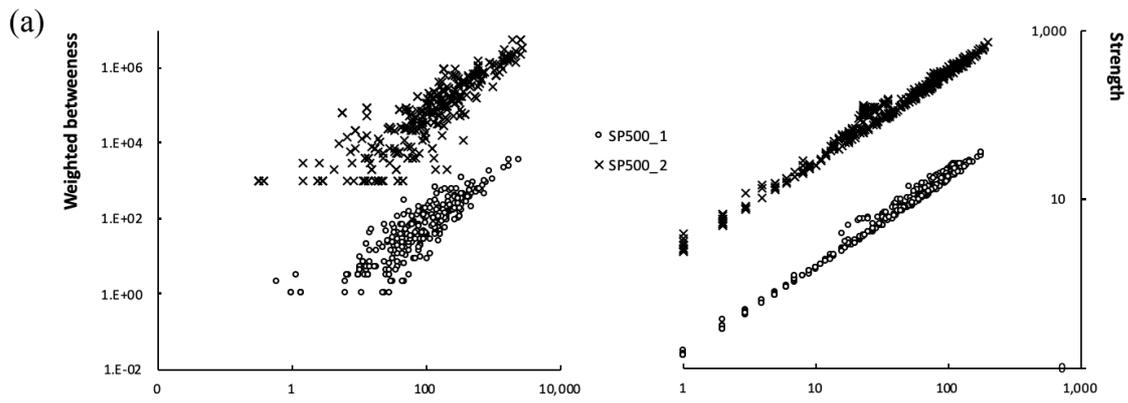

(a)

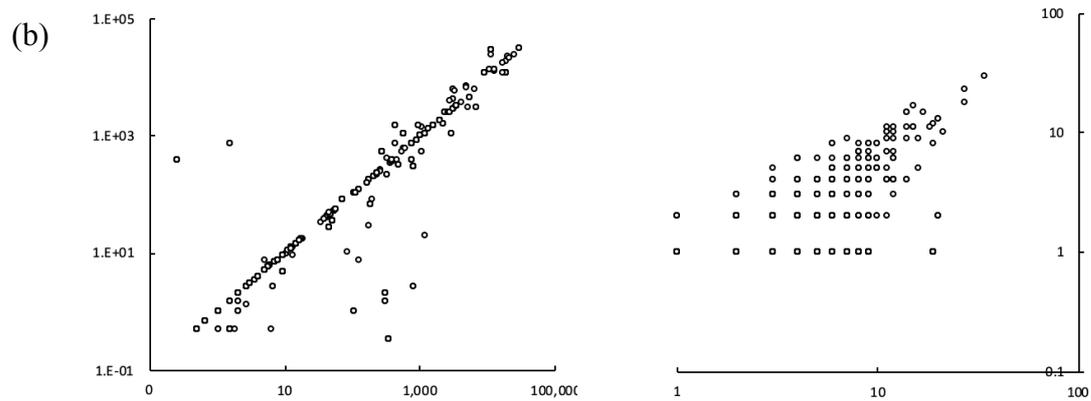

(b)

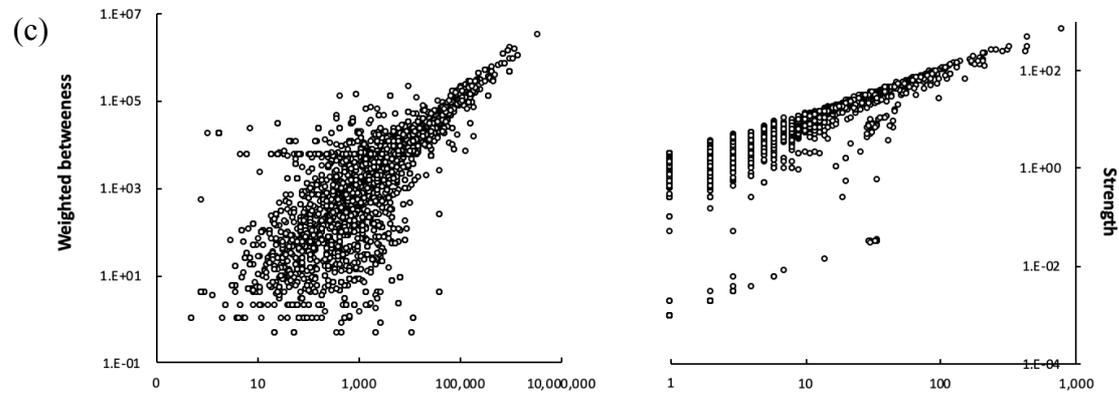

(c)

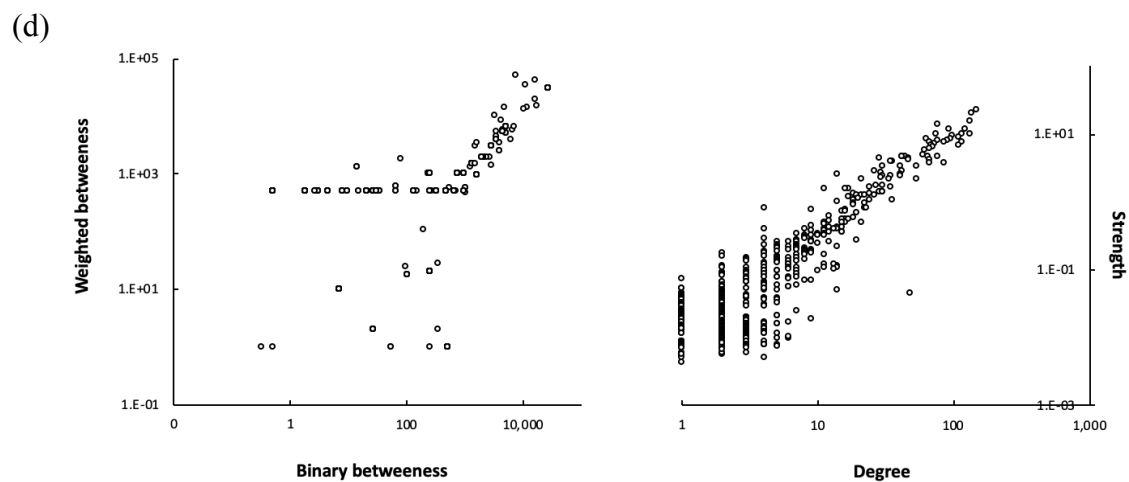

(d)

**Binary betweeness**           **Degree**

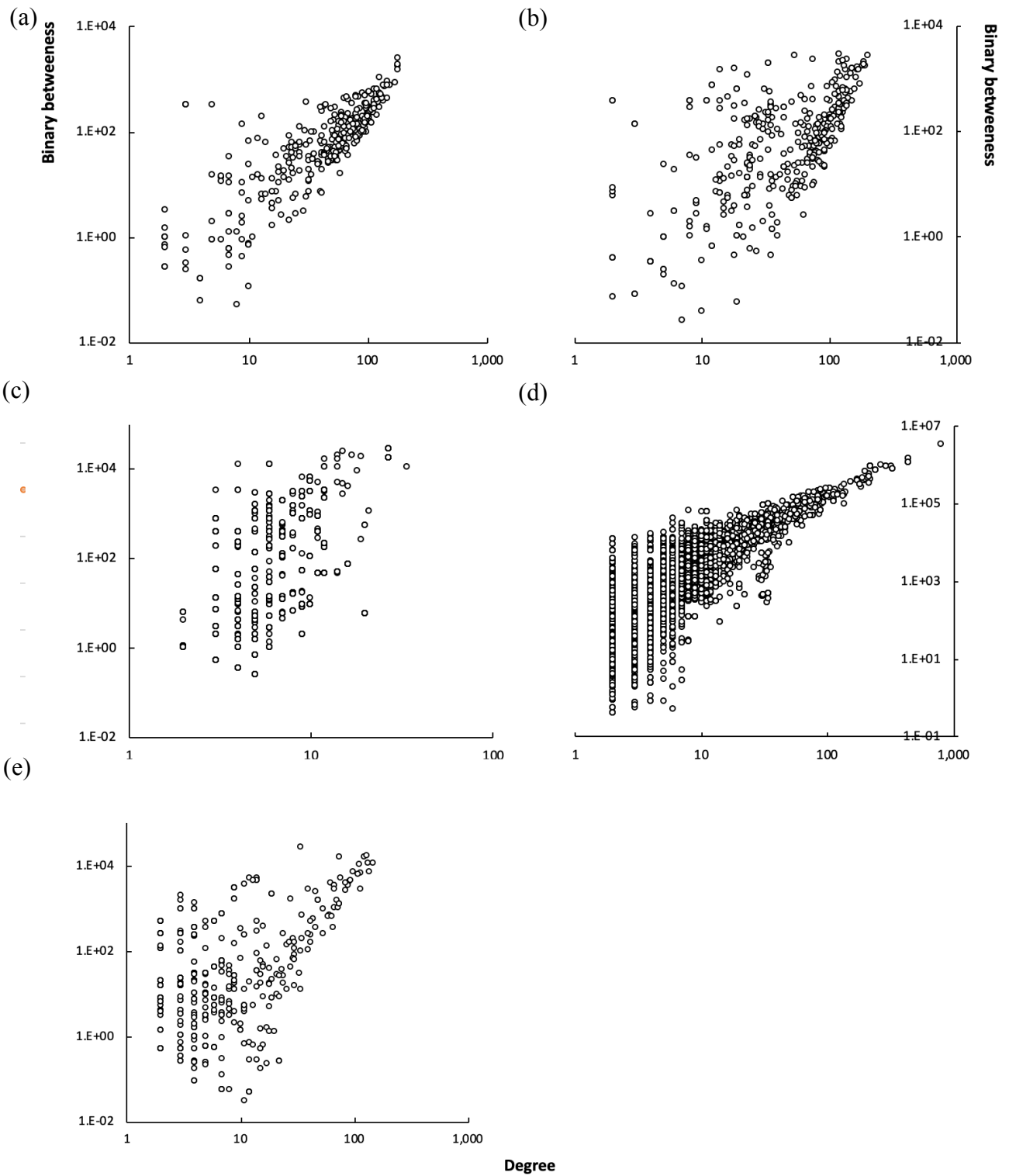

*Figure 9: Scattered plot of weighted betweenness and binary betweeness, Strength and degree for (a) two financial networks (with shifted scale for the SP500_2, (b) NetScience network , (c) Bitcoin network and (d) USAirport network*

*Figure 10: Scattered plot of binary betweenness and degree for (a) SP500_1 network, (b) SP500_2 network, (c) NetScience network, (d) Bitcoin network and e) Airport network*